\documentclass[lettersize,journal]{IEEEtran}
\usepackage{amsmath,amsfonts}
\usepackage{algorithmic}
\usepackage{algorithm}
\usepackage{array}
\usepackage[caption=false,font=normalsize,labelfont=sf,textfont=sf]{subfig}
\usepackage{textcomp}
\usepackage{stfloats}
\usepackage{url}
\usepackage{verbatim}
\usepackage{svg}
\usepackage{graphicx}
\usepackage{cite}
\begin{document}

\title{CCAT: Design, Implementation, and Testing of a System to Read Out over 10,000 280 GHz KIDs using RFSoC Electronics}

\author{Darshan A. Patel, Yuhan Wang, Cody J. Duell, Jason E. Austermann, James Beall, James R. Burgoyne, Scott Chapman, Steve K. Choi, Rodrigo G. Freundt, Eliza Gazda, Christopher Groppi, Zachary B. Huber, Johannes Hubmayr, Ben Keller, Lawrence T. Lin, Philip Mauskopf, Alicia Middleton, Michael D. Niemack, Cody Roberson, Adrian K. Sinclair, Ema Smith, Jeff van Lanen, Anna Vaskuri, Benjamin J. Vaughan, Eve M. Vavagiakis, Michael Vissers, Samantha Walker, Jordan Wheeler, Ruixuan(Matt) Xie
\thanks{D. A. Patel, Y. Wang, C. J. Duell, Z. B. Huber, B. Keller, L. T. Lin, A. Middleton, M. D. Niemack, E. Smith, B. J. Vaughan, E. M. Vavagiakis, and S. Walker are with the Department of Physics, Cornell University, Ithaca, NY 14853, USA. (e-mail: dp649@cornell.edu).

J. E. Austermann, J. Beall, J. Hubmayr, J. van Lanen, A. Vaskuri, M. Vissers, and J. Wheeler are with the Quantum Sensors Division, National Institute of Standards and Technology, Boulder, CO 80305, USA.

J. R. Burgoyne and R. Xie are with the Department of Physics and Astronomy, University of British Columbia, Vancouver, BC, Canada.

S. Chapman is with the Department of Physics and Atmospheric Science, Dalhousie University, Halifax, NS, Canada.

S. K. Choi and E. Gazda are with the Department of Physics and Astronomy, University of California, Riverside, CA 92521, USA.

Rodrigo G. Freundt and M. D. Niemack are with the Department of Astronomy, Cornell University, Ithaca, NY 14853, USA.

C. Groppi, P. Mauskopf, and C. Roberson are with the School of Earth and Space Exploration, Arizona State University, Tempe, AZ 85287, USA.

A. K. Sincair is with the NASA Goddard Space Flight Center, Greenbelt, MD 20771, USA.

E. M. Vavagiakis is with the Department of Physics, Duke University, Durham, NC 27710, USA.}
}

\markboth{IEEE Transactions on Applied Superconductivity, Special Issue for Low Temperature Detectors 2025, August~2026}%
{Shell \MakeLowercase{\textit{et al.}}: A Sample Article Using IEEEtran.cls for IEEE Journals}


\maketitle

\begin{abstract}
Over the past decade, kinetic inductance detectors (KIDs) have emerged as a viable superconducting technology for astrophysics at millimeter and submillimeter wavelengths. KIDs spanning 210 - 850 GHz across seven instrument modules will be deployed in the Prime-Cam instrument of CCAT Observatory's Fred Young Submillimeter Telescope at an elevation of 5600 m on Cerro Chajnantor in Chile's Atacama Desert. The natural frequency-division multiplexed readout of KIDs allows hundreds of detectors to be coupled to a single radio frequency (RF) transmission line, but requires sophisticated warm readout electronics. The FPGA-based Xilinx ZCU111 radio frequency system on chip (RFSoC) offers a promising and flexible solution to the challenge of warm readout. CCAT uses custom packaged RFSoCs to read out KIDs in the Prime-Cam instrument. Each RFSoC can simultaneously read out four RF channels with up to 1,000 detectors spanning a 512 MHz bandwidth  per channel using the current firmware. We use five RFSoCs to read out the \textgreater 10,000 KIDs in the broadband 280 GHz instrument module. Here, we describe and demonstrate the readout hardware, software and pipeline for the RFSoC system. We present a detector position map of the 280 GHz module focal plane and preliminary averaged spectral responses of a small subset of detectors from the TiN and first Al arrays. These measurements demonstrate our ability to simultaneously readout thousands of detectors, validate the end-to-end performance of the readout and optical systems, and represent a critical step toward reading out the $\sim$100,000 KIDs in Prime-Cam in its future full capacity configuration.

\end{abstract}

\section{Introduction}
\IEEEPARstart{T}{he} CCAT Observatory's six-meter aperture Fred Young Submillimeter Telescope (FYST) \cite{Niemack2016, ParshleyTelescope2018, Parshley2018} will provide a window into the millimeter and submillimeter sky in the 210 - 850 GHz band with an unprecedented number of detectors. Targeting a broad range of science goals \cite{CCAT2022, Stacey2018}, measurements with low atmospheric noise at these higher frequencies will be enabled by the location of the FYST at 5,600 m on Cerro Chajnantor in Chile's Atacama Desert due to the high elevation and dry climate \cite{Radford2016}. Prime-Cam \cite{Vavagiakis2018} is a first-generation science instrument of the FYST that can host up to seven instrument modules. Initial deployment in 2026 is planned with two broadband, polarization-sensitive modules centered at 280 GHz \cite{Vavagiakis2022, Choi2022, Duell2024, Vaskuri2025, Middleton2025} and 350 GHz. Prime-Cam instrument modules employ lumped element kinetic inductance detectors (KIDs). Lumped element KIDs are superconducting, thin film detectors with discrete capacitive and inductive components \cite{Doyle2008}, where a significant fraction of the inductance is derived from the kinetic inductance of the Cooper pairs. Incident photons with sufficient energy can break Cooper pairs in the inductor to alter the kinetic inductance, which produces a measurable shift in the resonant frequency and internal quality factor of the detector \cite{Day2003}. 
While the initial deployment of Prime-Cam will include $\sim$20,000 KIDs across the 280 GHz and 350 GHz instrument modules, the fully populated Prime-Cam will observe with $\sim$100,000 KIDs distributed across all seven modules. \par
\begin{figure}[h]
    \centering
    \includesvg[width=0.8\linewidth]{Figures/280GHz_focal_plane_rotated.svg}
    \caption{The full 280 GHz focal plane with one TiN KID array with Al feedhorns (bottom left) and two Al KID arrays with silicon-platelet feedhorns. Each array contains $\sim$3,400 polarization sensitive KIDs and is divided into six networks with $\sim$570 KIDs each.}
    \label{fig:focalplane}
\end{figure}

\begin{figure}[h]
    \centering
    \includesvg[width=0.95\linewidth]{Figures/det_map.svg}
    \caption{Map of 7,138 280 GHz detector positions extracted from near-field beam map measurements. Al network 1.6 and TiN network 1.3 are missing due to repairable issues with the low-noise amplifiers on those readout chains.}
    \label{fig:detmap}
\end{figure}
The three 280 GHz arrays shown in Fig. \ref{fig:focalplane} were fabricated by the Quantum Sensors Division at the National Institute of Standards and Technology (NIST) and have undergone extensive in-lab testing in the Prime-Cam and Mod-Cam \cite{Vavagiakis2022, Lin2025} receivers. 
The 280 GHz module is comprised of two silicon-platelet feedhorn-coupled \cite{Britton2010, Austermann2026} detector arrays with aluminium (Al) KIDs \cite{Vaskuri2025} and one Al feedhorn-coupled array with titanium-nitride (TiN) KIDs \cite{Austermann2018, Duell2024, Middleton2025} using a TiN/Ti/TiN trilayer\cite{Vissers2013}. Each array consists of $\sim1,700$ dual-polarization pixels, which amounts to $\sim3,400$ KIDs per array and over 10,000 KIDs for the full instrument module. Each array is further subdivided into six networks that are each read out using a RF feedline coupled to $\sim$570 KIDs. The networks are uniquely labeled in Fig. \ref{fig:focalplane} by detector type, array number, and network number (e.g., network three of the first Al array is labeled Al 1.3). \par

CCAT employs custom packaged, FPGA-based Xilinx ZCU111 radio frequency system on a chip (RFSoC) boards for the warm readout of the highly multiplexed KIDs. Previous work by \cite{Sinclair2022, Burgoyne2024, Sinclair2024} has presented the firmware and software design for the RFSoC readout system, laying the groundwork for this readout approach. In this work, we report on the current status and configuration of the RFSoC readout system, working toward its deployment in 2026, and demonstrate its use in-lab for instrument characterization.

\section{Readout Hardware and Software}\label{sec:software}

The KIDs are read out using Xilinx UltraScale+ ZCU111 RFSoC boards\footnote{\url{https://www.amd.com/en/products/adaptive-socs-and-fpgas/evaluation-boards/zcu111.html}} that are housed in custom 1U rack-mountable aluminum enclosures (Fig. \ref{fig:rfsoc-enclosure}) with additional electronics packaged at Arizona State University (ASU). For simplicity, in the remainder of this paper we refer to the custom packaged warm readout enclosure as the RFSoC and to the UltraScale+ ZCU111 itself as the RFSoC board. Inside the enclosures, four of the RFSoC board analog-to-digital converters (ADCs) and digital-to-analog converters (DACs) are broken out through a custom board and connected to digitally controlled variable attenuators providing up to 31.75 dB of attenuation. We refer to the attenuators connected to the DACs and ADCs as ``drive" and ``sense" attenuators respectively. An Opsero Ethernet FPGA Mezzanine card is attached to the RFSoC board to stream time-ordered data and synchronize the RFSoC board clock via Precision Time Protocol (PTP). The FPGA fabric is passively cooled using a copper heat sink thermally coupled to the enclosure. \par

\begin{figure}
    \centering
    \includesvg[width=0.8\linewidth]{Figures/RFSoC.svg}
    \caption{A radio frequency system on a chip (RFSoC) board (\textbf{A}) inside a custom 1U rack-mountable enclosure. The RFSoC FPGA fabric is cooled with a copper heat sink (\textbf{B}). The RFSoC analog-to-digital and digital-to-analog converters are broken out using a custom board and connected to digitally controlled variable attenuators (\textbf{C}) with up to 31.75 dB of attenuation. An Opsero Ethernet FPGA Mezzanine card (\textbf{D}) is attached to the board to stream time-ordered data. The remainder of the enclosure (\textbf{E}) contains the power supply, an unmanaged switch for breaking out Ethernet connections, and additional accessories for debugging.}
    \label{fig:rfsoc-enclosure}
\end{figure}

\begin{figure*}
    \centering
    \includegraphics[width=0.95\textwidth]{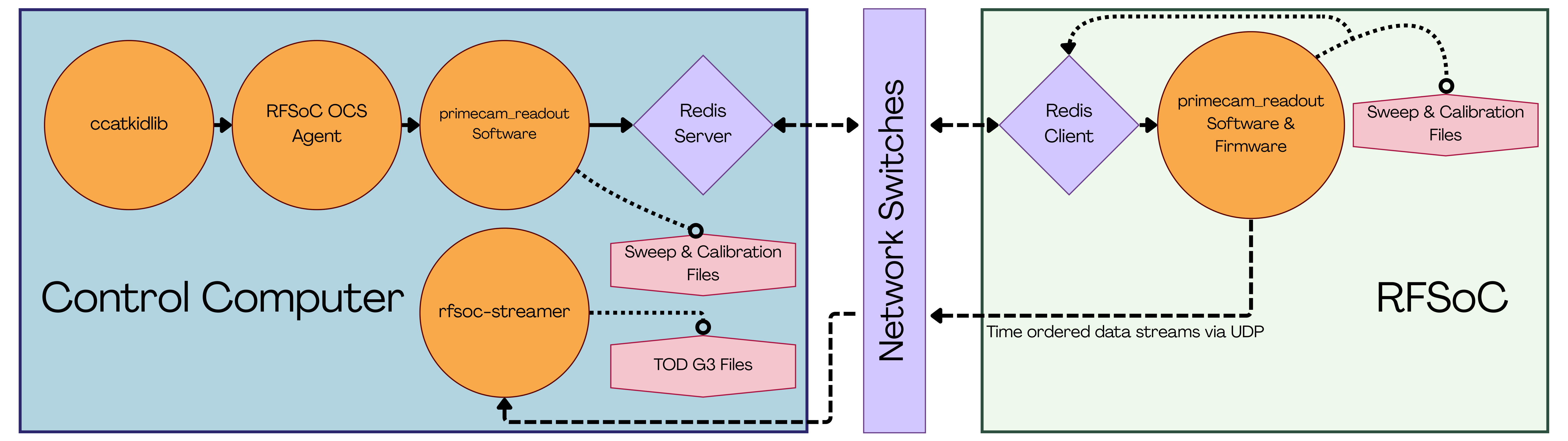}
    \caption{A high-level schematic diagram of the software pipeline used to take data with the RFSoCs from a centralized control computer. Only a single RFSoC is included in the diagram for simplicity but multiple can be connected at the network switch interface. Software packages are denoted by the orange circles with their hierarchy indicated by the solid arrows. Networking connections between the control computer and the RFSoCs are denoted in purple with the dashed arrows indicating communication via TCP (for the Redis server) or UDP (for time-ordered data streaming).  Data products are denoted by the red hexagons with the dotted lines indicating which software packages create them.}
    \label{fig:software-schematic}
\end{figure*}

The software used to collect data with the RFSoCs is summarized by the high-level schematic diagram in Fig. \ref{fig:software-schematic}. The RFSoCs are controlled using the \textit{primecam\_readout}\footnote{\url{https://github.com/TheJabur/primecam_readout}} firmware and low-level software \cite{Burgoyne2024}. The current firmware enables heterodyne measurements of the complex forward transmission $S_{21}$ of up to 1024 radio frequency waves, referred to as ``tones" from here on, per readout chain. Each RFSoC can simultaneously read out four chains for a total of up to 4096 tones. The tones are statically generated using a look-up table \cite{Sinclair2022} and can span a 512 MHz bandwidth around the numerically controlled local oscillator (NCLO) frequency. The \textit{primecam\_readout} software is installed on the RFSoC boards and on a central control computer. The control computer uses a Redis server to publish commands to the RFSoCs, which enables simultaneous communication with multiple boards \cite{Burgoyne2024}. The two main data products produced by \textit{primecam\_readout} are frequency sweeps and time-ordered data (TOD) streams. \par 

Since the tones are placed statically, the frequency sweeps are achieved by uniformly sweeping all of the tone frequencies via shifts in the NCLO frequency. The complex forward transmission of each tone is recorded at every frequency shift, and the data is saved as individual NumPy files per readout chain on the RFSoC. The data is also published to the Redis server so that it can be saved on the control computer. Publishing data through the Redis server is sufficient for the frequency sweeps but cannot be used to transmit TOD streams due to the much higher data volumes and required timing precision. Instead, TODs are packaged by the FPGA and streamed to the control computer using User Datagram Protocol (UDP) data packets \cite{Sinclair2022}. The core data stored in each packet are the complex transmission of each tone (separated into 32-bit in-phase $I$ and quadrature $Q$ components where $S_{21} = I + iQ$) and the timestamp at the end of data accumulation by the ADC, which is synced to the control computer using PTP. A sampling rate of $512\cdot10^6/1024^2\approx 488.28$ Hz is planned for in-field operation, but measurements with higher sampling rates can be taken. On the control computer, the UDP packets are captured using the Dockerized \textit{rfsoc-streamer}\footnote{\url{https://github.com/ccatobs/rfsoc-streamer}} software and packaged into the SPT-3G G3 file format\footnote{\url{https://github.com/CMB-S4/spt3g_software}}. A dedicated \textit{rfsoc-streamer} instance is employed for each RF chain to allow for scalability without data loss. The G3 files provide a compression factor of $\sim$2, and each file stores the data as segmented ``G3 frames'' of a user-specified length. The G3 frames allow for efficient loading of subsets of the data without requiring the full data set to be loaded into memory. \par
Together, the \textit{primecam\_readout} and \textit{rfsoc-streamer} software packages are sufficient for reading out KIDs with a RFSoC. The \textit{primecam\_readout} software, however, requires low-level knowledge of the KID readout scheme to properly initialize the RFSoC boards, generate tones, and take data. The low-level implementation accommodates flexibility in the readout but is not ideal for in-field operation during which a consistent, streamlined readout pipeline is preferable. The \textit{ccatkidlib}\footnote{\url{https://github.com/ccatobs/ccatkidlib}} software is under active development and facilitates higher level operation of the RFSoCs by combining the low-level \textit{primecam\_readout} commands into high-level commands corresponding to general KID tuning and data acquisition operations. To ensure reproducibility and aid in troubleshooting, relevant system state information about the control computer, RFSoCs, receiver, and detectors is saved in YAML files alongside every data file produced by \textit{ccatkidlib}. The \textit{primecam\_readout} commands are run by \textit{ccatkidlib} through a Simons Observatory (SO) Observatory Control System (OCS) agent \cite{Koopman2024}. Using an intermediary OCS agent allows \textit{primecam\_readout} to run in an isolated Docker container environment and allows access to the full suite of tools developed by SO for monitoring and logging the status of OCS agents. The OCS agent is also used to monitor the storage space of each RFSoC board and the temperatures of the FPGA fabric and ARM processor. 

\begin{figure*}[h]
    \centering
    \includesvg[width=0.9\linewidth]{Figures/primecam_readout.svg}
    \caption{\textit{Left:} Schematic diagram of the Prime-Cam and Mod-Cam cold readout chains adapted from Fig. 4.2 of \cite{DuellThesis}.  Attenuator values and distribution vary between individual networks and are still being finalized. \textit{Right:} Five rack-mounted radio frequency system on a chip (RFSoC) enclosures connected to the Prime-Cam RF readout harness with 36 SMA connections. Also pictured is a custom biasing system from ASU used to bias the 18 low-noise amplifiers.}
    \label{fig:readout_diagram}
\end{figure*}

\section{Readout Procedure}\label{sec:procedure}
The KID readout procedure with the RFSoC begins by searching for individual detector resonance dips. A frequency sweep with 1,000 equally spaced tones each sweeping 512 KHz is used to fully sample the available 512 MHz bandwidth, which we refer to as a ``VNA sweep". Each tone uses the same drive power but a randomized phase to reduce the crest factor of the overall waveform, which allows us to achieve higher power tones without over-driving the DAC \cite{Burgoyne2024}. The frequencies of detector candidates are identified using the SciPy \texttt{find\_peaks} peak finding algorithm \cite{Scipy2020} on the magnitude of the complex sweep data $|S_{21}| = \sqrt{I^2 + Q^2}$. Next, we take a frequency sweep using tones at the frequencies of the found $|S_{21}|$ resonance dips, which we refer to as a ``target sweep''. The target sweep data is then used to further refine the tone frequency placement with a number of methods under active exploration including placement at the resonant frequency, the frequency corresponding to the $|S_{21}|$ minima, or the frequency corresponding to the largest phase $\phi = \arctan{(Q/I)}$ vs. frequency gradient \cite{Gazda2026}. After refining the tone frequencies, the optimal tone drive powers are determined using target sweeps taken with different drive attenuations. Higher drive powers are preferred to reduce two-level system (TLS) noise \cite{Noroozian2009}, increase signal relative to amplifier noise, and optimize detector responsivity; however, KIDs exhibit bifurcation if driven with too high a power \cite{Swenson2013}. As it is difficult to operate KIDs in the bifurcated regime, we determine the attenuation required to place each KID roughly 1-3 dB below its bifurcation point using fits to the detector phase profiles \cite{Dai2022, Walker2024}. The optimal attenuation for each KID is converted to a tone drive power, and an overall network attenuation is chosen to place the total waveform amplitude in range of the DAC. A final target sweep is taken with the chosen network attenuation value, tone frequencies, and tone drive powers so that it can be used to calibrate the TODs and extract key resonator parameters such as resonant frequencies and quality factors. Finally, time-ordered $I$ and $Q$ data streams are taken with the tuned KIDs.

\section{Readout Demonstration}\label{sec:readout}

We use five RFSoCs connected to the Prime-Cam receiver as shown in the right panel of Fig. \ref{fig:readout_diagram} with the software packages described in Section \ref{sec:software} to read out the 280 GHz instrument module. 
Inside the receiver, the detector arrays are located at the 100 mK focal plane as shown in the left panel of Fig. \ref{fig:readout_diagram}. Flexible stripline RF circuits \cite{Keller2026} connect the 300 K readout feedthroughs to the 4 K stage. Coaxial cables then connect the 4 K stage to the 1 K stage and subsequently from the 1 K stage to the detector arrays. The exact distribution and amount of RF attenuation for each network is still being finalized to achieve optimal RF performance but is typically 20–30 dB in total, with the majority applied using cryogenic attenuators at the 4 K, 1 K, and 100 mK stages on the input chain. On the output chain, cryogenic RF amplification is provided by low-noise amplifiers (LNAs) developed at ASU and located at the 4 K stage. The LNAs are powered through individual DC chains in the receiver with the bias current supplied by a custom room-temperature biasing system produced by ASU and shown in the right panel of Fig. \ref{fig:readout_diagram}. 

\par

\begin{figure*}
    \centering
    \includesvg[width=0.9\textwidth]{Figures/det_dashboard.svg}
    \caption{The top panel shows the magnitude in ADC units of a VNA sweep of TiN network 1.1 overlaid with vertical lines at the frequencies of the 525 identified KIDs. Panels \textbf{A}, \textbf{B}, and \textbf{C} show target sweep data of a single detector taken over a 20 dB range of drive attenuations. The frequencies of the tones are indicated by the stars. Panel \textbf{A} is a polar plot of the target sweep IQ circles corrected and centered at the origin, and Panel \textbf{C} shows the phase vs. frequency response of the centered IQ circles. Panel \textbf{B} shows the magnitude of the target sweeps after removing a smooth baseline. 
    Panel \textbf{D} shows time-ordered data (TOD) taken at 4 dB drive attenuation in the IQ plane after it has been transformed using the target sweep data to be placed along the centered IQ circle. Also shown is the TOD rotated by ninety degrees to isolate the data in the dissipation direction. Panel \textbf{E} shows the time-ordered frequency and dissipation direction data in fractional frequency shift units, and Panel \textbf{F} shows the power spectral densities of the frequency and dissipation direction TODs.}
    \label{fig:detector-summary}
\end{figure*}

To demonstrate the readout procedure described in Section \ref{sec:procedure}, we present measurements of TiN network 1.1 in Fig. \ref{fig:detector-summary} encapsulating the full readout pipeline. The measurements were performed using the Prime-Cam receiver while optically open to the room and using a reduced aperture Lyot stop in the 280 GHz module to produce optical loading similar to that expected on-sky. The top panel of Fig. \ref{fig:detector-summary} shows the $|S_{21}|$ of a VNA sweep overlaid with vertical lines at the 525 frequencies identified by detector finding. We note that the shown $|S_{21}|$ is calculated directly from the raw ADC values and is therefore only proportional to the power transmitted through the readout chain. Panels \textbf{A}, \textbf{B}, and \textbf{C} in Fig. \ref{fig:detector-summary} show data of a single detector from target sweeps taken over a 20 dB range of drive attenuations. All target sweeps were taken with the tone placed at the frequency corresponding to the largest phase $\phi$ vs. frequency gradient as indicated by the stars. The raw target sweep $S_{21}$ data is corrected by multiplying by the factor $e^{i(2\pi \tau)f}$ where $f$ is the sweep frequency and $\tau$ is the cable delay obtained from the slope of the VNA sweep phase. The correction results in a circle in the IQ plane, which we refer to as the ``IQ circle", that is offset from the origin. We perform a least-squares fit of the IQ circle \cite{Chernov2005} and use the fitted center to transform the IQ circle to the origin \cite{Gao2008}. A polar plot of the IQ circles centered at the origin is shown in Panel \textbf{A} and the phase vs. frequency response of the centered IQ circles is shown in Panel \textbf{C}. The plots show the evolution of the detector towards bifurcation with decreasing drive attenuation/increasing drive power.  \par
Since the detector bifurcates at a drive attenuation of $\sim$ 2 dB, we take the TOD at a drive attenuation of 4 dB. The raw $S_{21}$ TOD is transformed in the same manner as the target sweep, which places the TOD along the centered IQ circle. In these coordinates, the phase and magnitude of the $S_{21}$ TOD are orthogonal components, corresponding respectively, to shifts in the frequency and dissipation direction of the detectors. The phase (frequency direction) TOD is converted to frequency units using an interpolating spline of the centered IQ circle phase profile. The frequency TOD is then converted to fractional frequency shift units using $x = \delta f/f = (f-f_0)/f_0$ where $f_0$ is the resonant frequency. The magnitude (dissipation direction) TOD is also converted to fractional frequency shift units by first rotating the TOD by ninety degrees as shown in Panel \textbf{D} of Fig. \ref{fig:detector-summary} and using the same interpolating spline on the phase of the rotated TOD. Panel \textbf{E} of Fig. \ref{fig:detector-summary} shows the time-ordered frequency and dissipation direction data in fractional frequency shift units, and Panel \textbf{F} shows the power spectral densities (PSDs) of those TODs. In this representation, photon and detector noise is predominantly displayed in the frequency direction, whereas readout noise is displayed equally in both directions. From the PSDs, the dissipation direction white noise is seen to be almost an order of magnitude smaller than that in the frequency direction, which indicates that readout noise is subdominant to photon and detector noise. Work is ongoing to demonstrate that the detector noise, primarily TLS noise, is subdominant to photon noise \cite{Mauskopf2014, Hubmayr2015} to achieve our desired noise hierarchy and sensitivity \cite{Choi2020}.  \par

\section{Optical Measurements}

We perform beam-map and spectral response measurements of the 280 GHz instrument module as part of its in-lab characterization and to demonstrate the readout capacity of the RFSoC. The beam-map and spectral response measurements were taken using the Prime-Cam and Mod-Cam receivers respectively. Both measurements were performed with the receivers optically open to the room, with a reduced aperture Lyot stop in the 280 GHz module, and with the full stack of optical components including silicon lenses, low-pass edge filters, and IR blockers \cite{Huber2024}. The KIDs were tuned as described in Section \ref{sec:procedure}, and all of the raw TODs are converted to fractional frequency shift units as described in Section \ref{sec:readout} before any measurement specific analysis.

\begin{figure}[h]
    \centering
    \includesvg[width=0.95\linewidth]{Figures/fts_beam_mapper.svg}
    \caption{\textit{Top Left:} A Fourier Transform Spectrometer (FTS) with a $\sim700$ K thermal blackbody source. The FTS is positioned in front of the Mod-Cam receiver while optically open to the room to measure the spectral response of the 280 GHz instrument module. An aperture stop is placed in front of the thermal source to limit the amount of light entering the FTS. \textit{Bottom Left:} An unobstructed view of the FTS showing the input and output mirrors, polarized wire grids, and motorized central mirror.  \textit{Right:} A vertically mounted Dover Motion motorized XY translation stage with an attached HawkEye IR-55 source used to measure near-field beam-maps of the 280 GHz instrument module. The IR source is surrounded by an activated charcoal cloth to reduce stray reflections. }
    \label{fig:fts_beam_mapper}
\end{figure}

For the beam-map measurement, we used a HawkEye IR-55\footnote{\url{http://www.hawkeyetechnologies.com/source-selection/pulsable/}} IR source attached to a vertically mounted Dover Motion motorized XY translation stage. The IR source was surrounded by an activated charcoal cloth\footnote{\url{https://buyactivatedcharcoal.com/activated-carbon-cloth-double-weave.html}} as shown in the right panel of Fig. \ref{fig:fts_beam_mapper} to reduce stray reflections. 
The IR source was modulated by a 10 Hz square wave with a 50\% duty cycle and 5 V peak-to-peak output voltage. A $15 \times 15$ point raster scan covering a $21 \times21$ cm area was performed with the XY stage positioned a few centimeters away from the receiver window. At each point in the scan, ten seconds of time-ordered data was taken with the KIDs, and the amplitude of the 10 Hz peak in the PSD of the fractional frequency shift TOD is extracted as the signal. The resultant near-field beam-map of each KID is fit to a symmetric, two-dimensional Gaussian to recover the approximate position of the KID in the focal plane. \par
A map of 7,138 KID positions is shown in Fig. \ref{fig:detmap} and is in good agreement with the expected geometry shown in Fig. \ref{fig:focalplane}. The two missing networks were due to repairable issues with the LNAs. Since the beam-maps were taken in the near-field, the measured KID beams were limited by the IR source output beam size, not diffraction. The large near-field beams and the 14 mm resolution of the beam-maps both contribute to the scatter seen in the detector map. Due to the limited translational range of the XY stage, Fig. \ref{fig:detmap} is the combination of three separate measurements, each centered on an individual array. Nevertheless, data was taken with all 16 of the functional networks for each of the three measurements, and these measurements demonstrate our ability to simultaneously read out thousands of KIDs with RFSoC electronics.\par
For the spectral response measurement, we used a polarizing Martin–Puplett interferometer in a Mach–Zehnder arrangement based on the PIXIE design \cite{Pan2019} as our Fourier-transform spectrometer (FTS).
The FTS was positioned in front of the receiver window as shown in the top left panel of Fig. \ref{fig:fts_beam_mapper} and aligned so that its output beam entered perpendicular to the window. 
A $\sim700$ K blackbody was used as a broadband source at one input of the FTS, while a room-temperature source was used at the other. A charcoal cloth covered aluminum block with a central hole was placed between the blackbody source and input mirror to act as an aperture stop.
Reflective surfaces outside of the FTS optical path were covered with the charcoal cloth to reduce stray light entering the receiver. One minute of time-ordered data was taken with the KIDs while continuously slewing the central mirror of the FTS at a speed of 1 mm/s over a maximum displacement of 2 cm from the center position. \par

\begin{figure}
    \centering
    \includegraphics[width=1\linewidth]{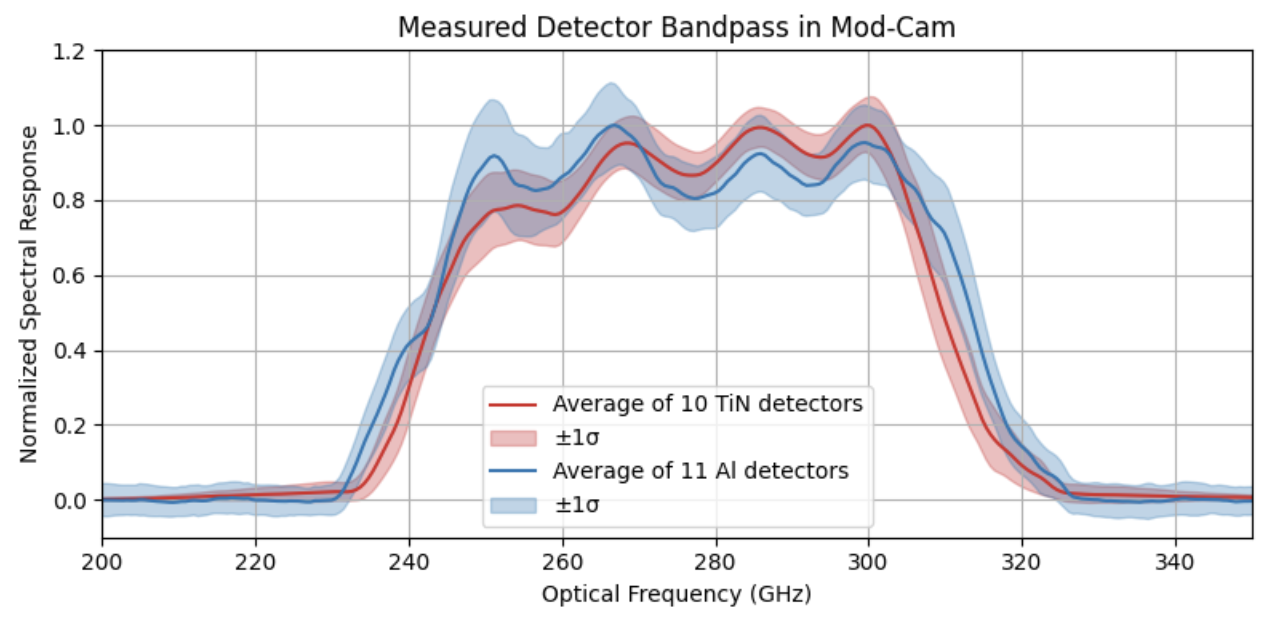}
    \caption{Preliminary averaged spectral response through the full Mod-Cam optical stack of a small subset of 280 GHz KIDs from TiN network 1.3 and Al network 1.2.}
    \label{fig:FTS}
\end{figure}

To calculate the spectral response, the position $y$ of the central mirror was logged over time and is used to convert the detector TOD times to the corresponding mirror positions. The mirror position is then converted to an optical path delay (OPD) $d$ using $d \approx 4y$ for small mirror displacements \cite{Pan2019}. The fractional frequency shift vs. OPD TOD is then Fourier transformed to produce the spectral response. Data was taken with all of the identified KIDs, but a cut down to the $\sim$10 highest signal-to-noise detectors was performed to only include the detectors fully illuminated by the FTS in the average. We expect the spectral response of the small subset of detectors to be a representative measurement since the spectral response depends primarily on the feedhorns (low-frequency cutoff) and filters (high-frequency cutoff), which should vary minimally across the focal plane. The preliminary averaged spectral responses in Fig. \ref{fig:FTS} show a $\sim80$ GHz bandpass centered at $\sim270$ GHz for both TiN network 1.3 and Al network 1.2, which aligns well with expectation \cite{Austermann2018}. 
We emphasize, however, that the measurements presented are a rough, initial characterization, and in-field measurements with Prime-Cam are planned with improved FTS alignment, the second Al array, and sampling of more regions of each array.

\section{Conclusion}
We discussed the firmware; \textit{primecam\_readout}, \textit{rfsoc\-streamer}, and \textit{ccatkidlib} software; and custom hardware used with the Xilinx UltraScale+ RFSoC ZCU111 to read out the 280 GHz instrument module KIDs in the Prime-Cam and Mod-Cam receivers. We described the KID readout and tuning procedure, which includes a full network VNA sweep, target sweeps around the identified detectors over a range of drive attenuations, and time-ordered \textit{I} and \textit{Q} data streams. We demonstrated the readout pipeline in Prime-Cam using TiN network 1.1. \par

We validated the end-to-end performance of our readout and optical systems with spectral response and beam-map measurements. The preliminary averaged spectral responses of a small subset of KIDs from TiN network 1.3 and Al network 1.2 showed $\sim$80 GHz bandpasses centered at $\sim$270 GHz. The near-field beam-maps were used to construct a map of 7,138 detector positions in the focal-plane, demonstrating our ability to simultaneously read out thousands of detectors with RFSoC electronics. Improved beam-map measurements using planet scans are planned for in-field operation, and in-field spectral response measurements with Prime-Cam are planned with the second Al array and more accurate FTS alignment.

\section*{Acknowledgments}
The CCAT project, FYST and Prime-Cam instrument have been supported by generous contributions from the Fred M. Young, Jr. Charitable Trust, Cornell University, and the Canada Foundation for Innovation and the Provinces of Ontario, Alberta, and British Columbia. The construction of the FYST telescope was supported by the Gro{\ss}ger{\"a}te-Programm of the German Science Foundation (Deutsche Forschungsgemeinschaft, DFG) under grant INST 216/733-1 FUGG, as well as funding from Universit{\"a}t zu K{\"o}ln, Universit{\"a}t Bonn and the Max Planck Institut f{\"u}r Astrophysik, Garching. The construction of the 350 GHz instrument module for Prime-Cam is supported by NSF grant AST-2117631. The completion and deployment of the Prime-Cam instrument with the initial instrument modules is supported by a generous contribution from Alex Gerko, Founder and CEO of XTX Markets. S. Walker acknowledges support from the National Science Foundation under Award No. 2503181. S. Chapman acknowledges both the Natural Sciences and Engineering Research Council of Canada and the Canadian Foundation for Innovation.

\bibliographystyle{ieeetr}
\bibliography{LTD}

\end{document}